# Deep Transfer Learning for Texture Classification in Colorectal Cancer Histology

Srinath Jayachandran, Ashlin Ghosh

Robert Bosch, Bangalore, India
{jayachandran.srinath,ghosh.ashlin}@in.bosch.com

**Abstract.** Microscopic examination of tissues or histopathology is one of the diagnostic procedures for detecting colorectal cancer. The pathologist involved in such an examination usually identifies tissue type based on texture analysis, especially focusing on tumour-stroma ratio. In this work, we automate the task of tissue classification within colorectal cancer histology samples using deep transfer learning. We use discriminative fine-tuning with one-cycle-policy and apply structure-preserving colour normalization to boost our results. We also provide visual explanations of the deep neural network's decision on texture classification. With achieving state-of-the-art test accuracy of 96.2% we also embark on using deployment friendly architecture called SqueezeNet for memory-limited hardware.

## 1 Introduction

According to the statistics provided by the American Cancer Society, colorectal cancer (CRC) is the third and second most commonly occurring cancer in men and women, respectively [1]. Histopathology provides one of the diagnosis procedures wherein, suspicious tissue is sampled by biopsy and examined under a microscope. A typical pathology report consists of tissue cell structural information which is used by the pathologist to decide any presence of malignant tumours. Such histological samples may typically contain more than two tissue types. Automating texture classification in CRC histology images will aid pathologists in making informed clinical decisions. Figure 1 represents randomly sampled histological images of eight different tissue types in CRC. Histology image analysis for cancer diagnosis can be extremely challenging because of issues with slide preparation, variations in staining and inherent biological structure [2]. This makes the need for domain-specific input very important for feature generation. Deep learning being domain agnostic can make use of rich information present in histology images.

## 2 Related work

Studies relating to different texture analysis in human CRC has been lacking, although [2] has introduced a similar dataset containing 8 different classes of textures in CRC. They used traditional machine learning with hand-engineered texture descriptor features like histogram, local binary patterns, gray level co-occurrence matrix, gabor filters and perception like features with a reported accuracy of 87.4%. Work involving the use of deep neural networks on this dataset is very limited. A neural network designed from scratch on a small dataset will lack generality and perform poorly on unseen data. Authors in [3] have designed a fully convolutional neural network (CNN) of 11 layers which could only classify with 75.5% accuracy. In another work [4], stain decomposition has boosted the performance on this dataset. They have derived hematoxylin and eosin (H&E) image components using an orthonormal transformation of the original RGB images and fed it into a bilinear convolutional neural network giving an accuracy of 92.6%. Since transfer learning based classification approach has not been taken on this dataset we introduce methodologies that achieve state-of-the-art performance.

## 3  Methodology

### 3.1  Dataset and Preprocessing

The CRC dataset contains 5000 RGB histological images of 150*150 px each belonging to one of the 8 tissue categories. Each category has 625 images of H&E stained tissue samples digitized with an Aperio ScanScope [2]. We follow two steps of image preprocessing for this dataset. Firstly, the H&E staining process enables a clear view of morphological changes within a tissue [5].

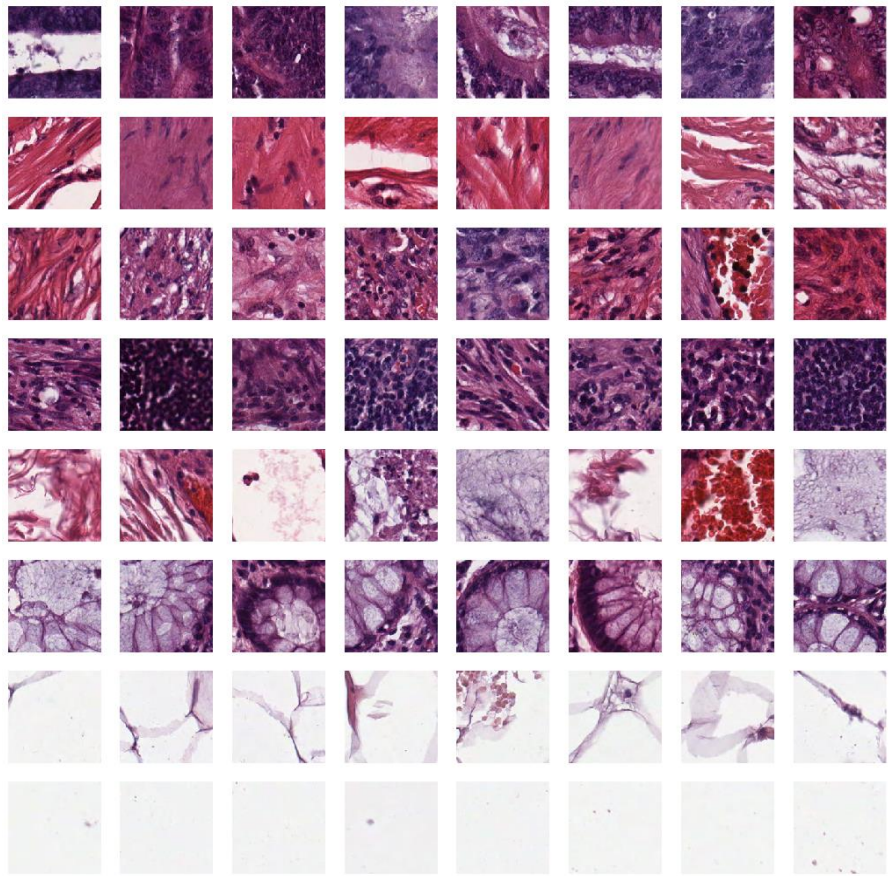

**Fig. 1.** Randomly sampled 8 images from each class (row) - a) Tumour epithelium b) Simple stroma c) Complex stroma d) Immune cell e) Debris f) Mucosal glands g) Adipose tissue h) Background (no tissue) in row-wise order starting from the top.

But this process is prone to undesirable colour variations across tissue types because of differences in tissue preparation, staining protocols, the colour response of scanners, and raw materials used in stain manufacturing. Any learning algorithm weighing-in more on the colour variation will lead to error in classification [6]. We use a structure-preserving colour normalization technique with sparse stain separation on these images given by [7]. Keeping one target image, other images are normalized by combining their respective stain density maps with a stain colour basis of the target image thus preserving the morphology. Figure 2 illustrates the effect of colour normalization. Second, the colour normalized dataset is then scaled between 0 and 1 followed by normalizing each channel to the ImageNet dataset [8].

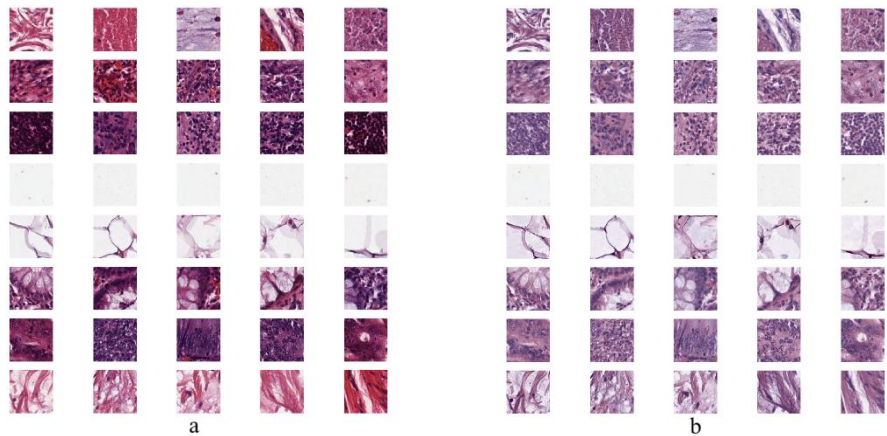

**Fig. 2.** Effect of structure-preserving colour normalization on a) Raw images as illustrated in b) Structure preserved colour normalized images.

### 3.2 Data Augmentation

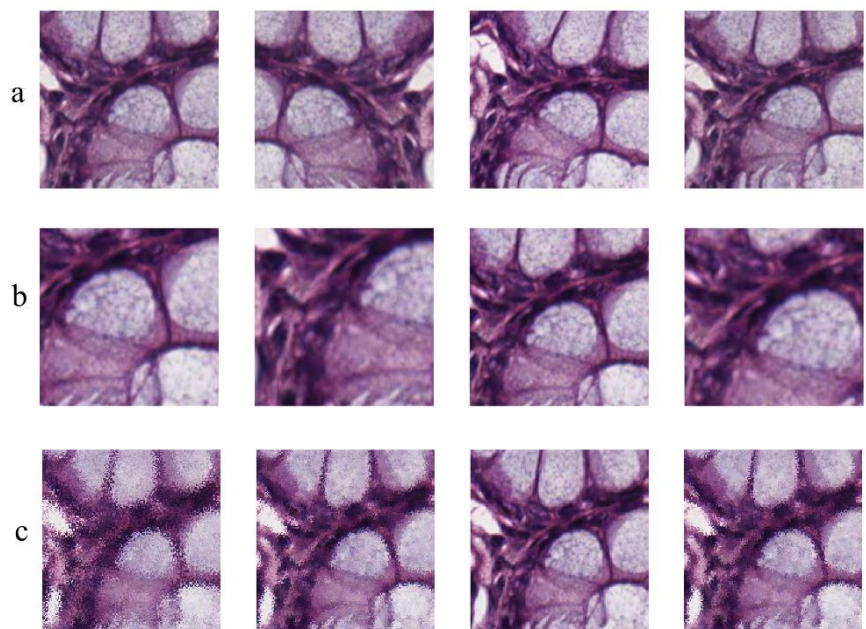

**Fig. 3.** Examples of image transformations for data augmentation using a) Rotations b) Random zoom crops and c) Jitter

With limited data, convolutional neural networks may overfit. Data augmentation improves the generalization capability of these networks by transforming images such that the network becomes robust to unseen data [9]. Random zoom crops were applied, as image patches in histopathology are invariant to translation in the input space. Tissue diagnosis is rotation invariant, which means that the pathologists can study histopathological images from different orientations. We introduce vertical and horizontal flips, and rotations, restricted to 90, 180 and 270 degrees because of interpolation issues. The other augmentation techniques used were lighting, warps, gaussian blur, and elastic deformation. We applied in-memory dynamic data augmentation that applies random transformations on a batch of images during training. Figure 3 shows examples of a few transformations.

### 3.3 Transfer Learning using SqueezeNet

Advancements in Deep Learning has led to super-human level performance on ImageNet large scale visual recognition challenge. State-of-the-art deep neural networks (DNN) trained on ImageNet dataset possess generic feature computation capabilities like gabor filters and colour bobs in their first layer that are very generic to any dataset or task. Whereas the final layer of these architectures become task-specific [10]. Given a new target visual dataset with a limited number of training examples, features from the pre-trained neural networks can be repurposed to adapt to this new dataset, called transfer learning. Since the AlexNet [11] breakthrough in ImageNet classification, many variants of convolutional neural networks have been submitted to the ImageNet challenge achieving state-of-the-art results. There is a high correlation between top-1 accuracy ImageNet architectures and their transfer learning capabilities [12], which makes it obvious to pick an architecture that performs the best on ImageNet. The focus of the majority of the models has not been on resource utilization hence are not practically deployable on resource-limited hardware. In this work, we choose a state of the art DNN architecture in terms of speed and accuracy tradeoff namely, SqueezeNet [13]. Squeezenet is a convolutional neural network that is carefully designed such that it has few parameters but with competitive accuracy on ImageNet. For this, they follow strategies like replacing 3x3 filters with 1x1 filters, reduce the number of input channels to 3x3 filters and maintaining large activation maps for the convolutional layers by downsampling late in the network. These strategies are bundled into a module called Fire module (Figure 4) which consists of a set of 1x1 filters in the squeeze convolutional layer, and a mix of 1x1 and 3x3 filters in the expand layer. With only 1,267,400 parameters and model size of 4.85 MB, SqueezeNet turns out to be a very lightweight model. The network macro architecture and architectural dimensions are presented in Figure 5 and Table 1, respectively.

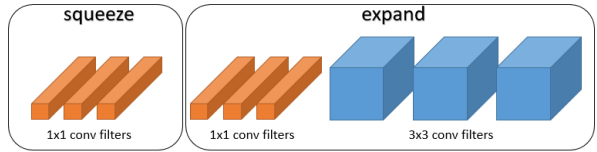

**Fig. 4.** Fire module of SqueezeNet

| Learning Rate | Layer Group | Layer | #1x1 | #1x1 | #3x3 | Filter size/ Stride | Output shape |
|---|---|---|---|---|---|---|---|
| | | Input image | | | | | 224x224x3 |
| 0.0001 | 1 | Conv1 | | | | 7x7 / 2 | 96x109x 109 |
| | | Maxpool1 | | | | 3x3 / 2 | 96x54x 54 |
| | | Fire1 | 16 | 64 | 64 | | 64x 54x 54 |
| | | Fire2 | 16 | 64 | 64 | | 64x 54x 54 |
| 0.003 | 2 | Fire3 | 32 | 128 | 128 | | 128x 54x 54 |
| | | Maxpool2 | | | | 3x3/2 | 256x 27x 27 |
| | | Fire4 | 32 | 128 | 128 | | 128x 27x 27 |
| 0.006 | 3 | Fire5 | 48 | 192 | 192 | | 192x 27x 27 |
| | | Fire6 | 48 | 192 | 192 | | 192x 27x 27 |
| | | Fire7 | 64 | 256 | 256 | | 256x 27x 27 |
| | | Maxpool3 | | | | 3x3/2 | 512x 13x 13 |
| | | Fire8 | 64 | 256 | 256 | | 256x 13x 13 |
| 0.01 | 4 | Adaptiveavgpool2d | | | | | 512x 1x 1 |
| | | Adaptivemaxpool2d | | | | | 512x1x 1 |
| | | Flatten | | | | | 1024 |
| | | BatchNorm1d | | | | | 1024 |
| | | Dropout | | | | | 1024 |
| | | Linear | | | | | 512 |
| | | Batchnorm1d | | | | | 512 |
| | | Dropout | | | | | 512 |
| | | Softmax | | | | | 8 |

**Table 1.** SqueezeNet architectural dimensions

Pretrained SqueezeNet was used as a backbone network and its penultimate layer is used as a feature extractor. The final output layer is replaced with a series of fully connected layers with Kaiming initialization [14], coupled with BatchNorm [15], and Dropout [16] layers which we call the head (layer group 4, Table 1). The rectified linear unit (ReLU) was used as the activation function.

### 3.4 Finding the optimal learning rate for superconvergence

The ability of an architecture to converge towards global minima on the loss function topology is an active area of research and is guided by hyperparameters like learning rate, batch size, momentum, and weight decay. Optimizers like Adam, AdaGrad, AdaDelta, and Nesterov momentum use piecewise constant learning rate, starting with a global learning rate while carefully reducing it on test set reaching a plateau [17]. Such strategies do not have a mechanism to automatically choose large learning rates that may help the network converge faster. We unfreeze the weights of the head (layer group 4, Table 1) and make them learnable while freezing the rest of the layers and fine-tune for two epochs. Next, we unfreeze the entire network and test its ability to superconverge [18].

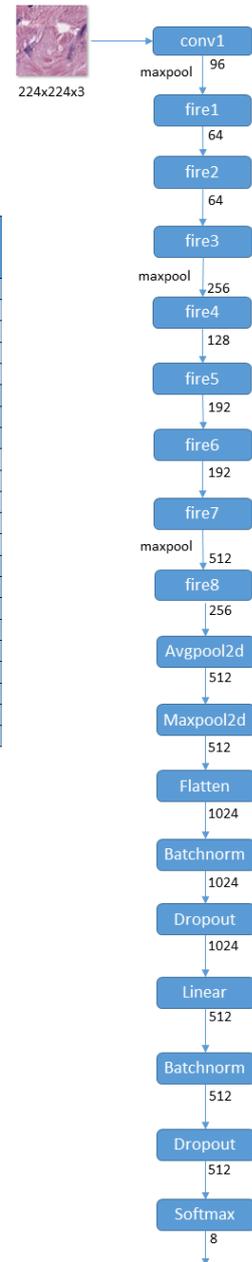

**Fig. 5.** SqueezeNet Macro architecture

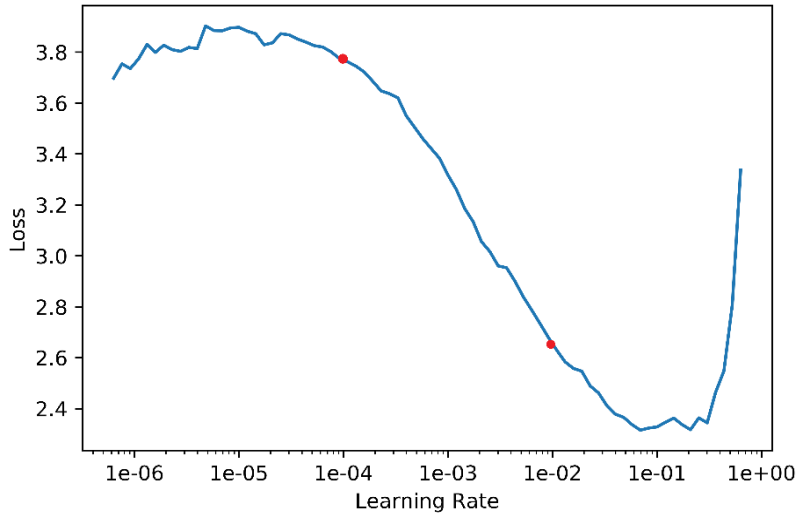

**Fig. 6.** Learning Rate Range Test – Training the entire network over a range of learning rates. The red markers define boundary values for a learning rate range where the network is still learning.

We run a learning rate (LR) range test [19], a mock training on the network on a large range of learning rates for 100 batches and generate a loss vs learning curve as given in Figure 6. This gives us an idea of the maximum learning rate ($L_{max}$) up to which the model converges, beyond that the test or validation loss starts increasing leading to overfitting and poor accuracy. Learning rates between 0.0001 and 0.01 prove to reduce the loss, whereas beyond 0.01 the network starts to unlearn.

### 3.5 Discriminative fine-tuning with One-cycle-policy

Instead of using a global learning rate and monotonically decreasing it, we implement one cycle policy (cyclical learning rate) [20] with decoupled weight decay (AdamW – beta1 = 0.9 and beta2 = 0.99) [21] that lets the learning rate cyclically vary between reasonable boundary values. This lets the network to converge faster and attain improved accuracy. Learning rate slightly lesser than $L_{max}$ as the maximum bound, and 10 times less to the maximum bound as the lower bound we train the network by starting at the lower bound and linearly increasing the learning rate up to maximum bound. At the same time momentum is decreased from 0.95 to 0.85 linearly. Then we perform cosine-annealing on learning rate up to 0 while applying symmetric cosine annealing on momentum from 0.85 to 0.95 as shown in Figure 7 [22].

Pretrained architectures exhibit different levels of information in their layers, starting from initial layers learning generic features to the final layer learning task-specific high-level features. Hence, different layers require different learning rates when being fine-tuned for a new task [23]. As shown in Table 1, the network is divided into 4 layer groups. The maximum bound learning rate discovered using the LR range test is assigned to the final (4th) layer group and the preceding layers are assigned with evenly spaced decreasing learning rates up to boundary value marked by red. Each layer then undergoes one-cycle-policy with new maximum and minimum bounds.

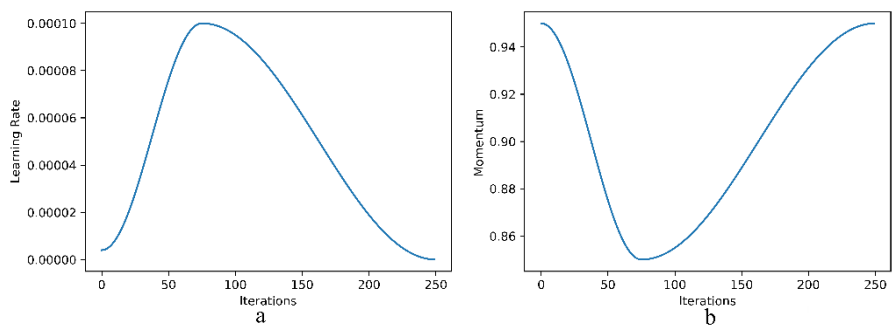

**Fig. 7.** Progression of a) learning rate and b) momentum during one cycle training policy.

## 4 Visual explanation using Gradient-based localization

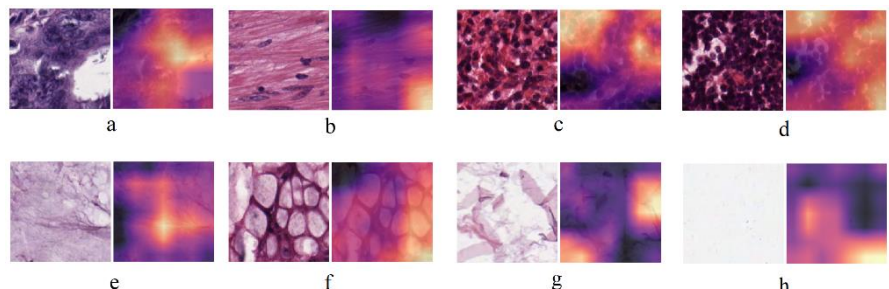

**Fig. 8.** Raw image vs Predicted heatmap. a) Tumour epithelium b) Simple stroma c) Complex stroma d) Immune cell e) Debris f) Mucosal glands g) Adipose tissue h) Background (no tissue). The brighter parts in the heatmap represent network's attention leading to correct prediction.

Making deep learning models transparent and explainable helps in understanding the failure modes as well as establishing trust and confidence in its users. But decomposing deep neural networks into intuitive and interpretable components is difficult. A technique known as Class Activation Map (CAM) is very popular for interpreting the decisions made by deep learning models [24] but is limited to architectures with feature maps directly preceding the softmax layers. Hence, we provide visual explanations of texture detection by the networks using a generalization of CAM known as Grad-CAM [25]. Similar to CAM, Grad-CAM generates a weighted combination of feature maps but followed by a ReLU. Grad-CAM extracts gradients from a CNN's final convolutional layer and uses this information to highlight regions most responsible for the prediction.

The images in Figure 8 demonstrate the original image of the tissue and the same image with a superimposed attention map created using Grad-CAM. Such visualizations could be very much useful in medical diagnosis since it reflects which parts of the tissue is affecting the model's predictions most. Possibly, this information can guide the practitioner in CRC biopsies to confirm the suspected diagnosis.

## 5 Results

The dataset is shuffled and randomly stratified sampled into 3 sets, train, validation and test of 60:20:20. We conducted different experiments related to the use of pre-trained network SqueezeNet and optimizer with one cycle policy. In Table 2 we compare the results of Squeezenet architecture trained with and without the pre-trained weights. Also, the difference in the results is analyzed when traditional piecewise constant learning rate scheduler optimizer like Adam is used.

**Table 2.** Comparison of one cycle policy with Adam optimizer

| Squeezenet Pretrained weights | Optimizer | Validation set(%) | Test set(%) |
| --- | --- | --- | --- |
| True | One cycle policy with AdamW | 97.4 | 96.2 |
| False | One cycle policy with AdamW | 80.1 | 75.6 |
| True | Adam | 71.5 | 64.8 |

It is realized that while training the network with Adam [17] and pre-trained weights as true, the network achieves below par accuracy. One cycle policy and AdamW without using the pre-trained weights performs better than Adam, and with weights, we achieve the state of the art results of 97.4% and 96.2% on the validation and test set respectively. These experiments were carried out using fastai [26] and Tensorflow [27] frameworks. We computed the receiver operating characteristic (ROC) curves (Figure 9) for each of the classes based on different threshold settings and generated the area under the curve (AUC) plots. The ROC curve is a plot of the true positive rate (TPR) against the false positive rate (FPR) at various threshold settings. An AUC of 1 is a perfect scenario of the model predicting every class correctly in the test set. With the SqueezeNet architecture, by computing average ROC the overall sensitivity and specificity is approximately 99%.

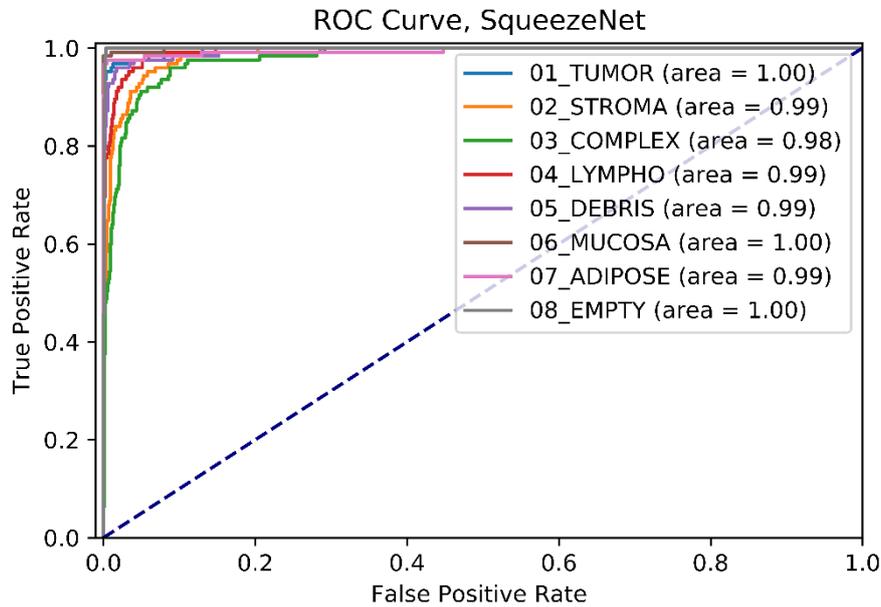

**Fig. 9.** ROC curves for the test set using SqueezeNet's fine-tuned architecture.

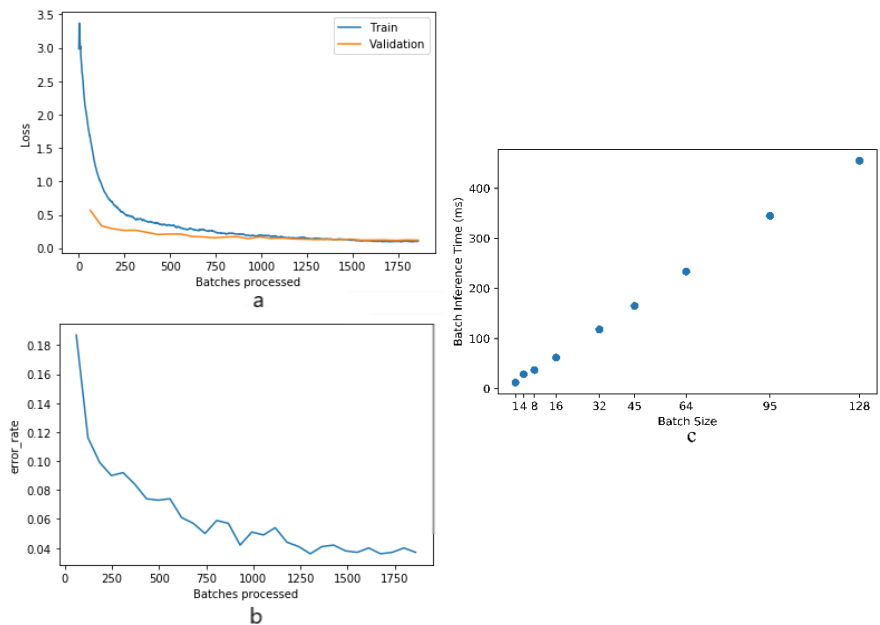

**Fig. 10.** a) Train and validation loss, b) error rate, and c) batch inference time vs batch size

Figure 10 represents 14 epochs of converging a) training and validation loss curves and b) accuracy (1-error rate) curve, when trained with a batch size of 32 and one threshold setting. The network saturated at an accuracy of 97.4% on the validation set, giving 96.2% on the test set.

The final model is serialized and made ready for deployment. We tested the model's performance in terms of inference time vs batch size (Figure 10, c) on a 640 cores Quadro M2000M GPU.

## 6      Conclusion

In this paper, we achieve state-of-the-art results in texture classification in human colorectal cancer using transfer learning with superconvergence. The work also takes into consideration deployment friendly DL model and network visualization to make neural networks decision making more transparent and explainable. Squeezenet, a model of very small size (4.8 MB) is used to demonstrate the results. Various data augmentation techniques and structure preserving colour normalization were also used to boost the results. For future work, we aim to investigate tumour progression from the learned network using a similar dataset.